# *Ab initio* approach to the elastic, electronic, and optical properties of MoTe$_2$: A topological Weyl semimetal

B. Rahman Rano[1], Ishtiaque M. Syed[1], S. H. Naqib[2,*]

[1]Department of Physics, University of Dhaka, Dhaka-1000, Bangladesh

[2]Department of Physics, University of Rajshahi, Rajshahi-6205, Bangladesh

*Corresponding author e-mail: salehnaqib@yahoo.com

**ABSTRACT:**

The topological Weyl semimetal MoTe$_2$, in the orthorhombic phase, is an important system both from the point of view of fundamental physics and potential applications. In this study we have investigated the elastic, electronic, bonding and optical properties of MoTe$_2$ in detail using density functional theory within both local density and generalized gradient approximations. Study of the elastic constants and moduli indicates that MoTe$_2$ is a relatively soft material with high level of machinability. Mechanical stability conditions are fulfilled. The compound possesses elastic and mechanical anisotropy and is prone to brittle fracture. Elastic parameters indicate that both covalent and metallic bondings are present in MoTe$_2$. This is supported by the charge density distribution mapping and Mulliken and Hirshfeld bond population analyses. Debye temperature, $\theta_D$, has been calculated. A relatively low value of Debye temperature also supports the scenario where bonding strengths are weak. The bulk electronic band structure calculations reveal clear indications of semi-metallic character. A pseudogap in the electronic energy density of states at the Fermi level indicates high level of electronic stability. Features reminiscent of the Dirac cone is observed close to the Fermi level. There is significant electronic anisotropy. Bands running in the crystallographic *c*-direction are non-dispersive with high carrier effective mass. Investigation of optical constants demonstrate that MoTe$_2$ possess excellent reflecting characteristics over a wide spectral range encompassing the infrared to ultraviolet regions. The compound also has high refractive index in the visible range. MoTe$_2$ is optically anisotropic, reflecting the anisotropic nature of the electronic band structure. The energy dependent optical parameters show metallic features and are in complete accord with the electronic density of states calculations.

# I. Introduction

Weyl fermion was first proposed in 1929 [1] but these Weyl fermion semimetals were experimentally discovered recently in the TaAs family of compounds [2,3]. Shortly afterwards type-II Weyl semimetal (WSM) was predicted in $WTe_2$ [4]. Type-II WSMs have tilted Weyl cones and they violate Lorentz symmetry. $T_d$-$MoTe_2$ is a sister compound of $WTe_2$, both crystallizing in orthorhombic lattice. At room temperature semi-metallic $MoTe_2$ has monoclinic structure. A structural phase transition from monoclinic to orthorhombic occurs at around 250 K [5]. In this study we have focused on the orthorhombic $T_d$ phase of $MoTe_2$ and from now on we will simply refer it to as $MoTe_2$.

The structural and electronic properties of $MoTe_2$ were studied in 1987 [6]. $MoTe_2$ was first predicted, from the Fermi surface and band structure calculations, to be a type-II Weyl semimetal [7,8] followed by experimental evidence of Fermi arcs [9–11]. Fermi arcs are formed between Weyl points (monopoles with fixed chirality) [12]. Superconductivity was found in $MoTe_2$ with a transition temperature of 0.10 K at ambient pressure and of 8.2 K at 11.7 GPa [13]. Later it was found that the crystal structure can be tuned mechanically and hence superconductivity is decoupled from the structural transition [14]. Due to its enormous possibilities $MoTe_2$ drew a lot of research interest over the past couple of years. Anticorrelation between polar lattice instability and superconductivity was also found in $MoTe_2$ [15]. Berry curvature dipole, surface superconductivity, Barkhausen effect, large magnetoresistance and planar Hall effect were observed by different groups [16–20]. Very recently some optical properties of $MoTe_2$ were studied [21]. The optical conductivity and reflectivity spectra were obtained experimentally in their work.

As far as we are aware of, the mechanical properties of $MoTe_2$ have not been studied in any detail. A thorough theoretical understanding of optical properties is still lacking. It should be stressed that comprehensive understanding of mechanical and optical properties are needed to unlock the potential of a material for possible applications. In this paper we have concentrated on investigation of the elastic and optical properties of the titled material supported by the band structure and electronic energy density of states. Electronic band structure and optical parameter studies show the semi-metallic characteristics of $MoTe_2$. The elastic constants reveal the soft nature of the compound. The optical parameters' spectra tell us that the plasma frequency should be around 16 eV. The reflectivity spectra over a broad range of frequency and high refractive index in the visible range assert that $MoTe_2$ is a good candidate for optoelectronics applications. High degree of machinability indicates that this compound holds promise for engineering applications.

The rest of this paper has been structured as follows: In Section II, we briefly discussed the computational methodology and crystal structure. In Section III, we have presented the results of our computations and analysed them. In this part we have discussed the elastic properties, Debye temperature, electronic properties, charge density distribution, population analysis and optical properties under different subsections. Finally, the important features of our calculations are summarized in Section IV.

**II. Methods and crystal structure**

The most popular approach to *ab initio* modelling of structural and electronic properties of solids is DFT. Here the ground state of the crystalline system is found by solving the Kohn-Sham equation [22]. In this study, we have used local density approximation (LDA) and generalized gradient approximation (GGA) schema as contained within the Cambridge Serial Total Energy Package (CASTEP) [23] code designed to implement DFT based calculations. It is known that, in general, LDA contracts the lattice due to localised nature of the trial orbitals whereas GGA overestimates the lattice constants. Comparing with the experimentally measured lattice parameters, it was found that LDA gives better estimates of the ground state structural parameters for $MoTe_2$. Therefore, we have presented the results of LDA calculations in the succeeding sections.

Vanderbilt-type ultra-soft pseudopotentials were used to take into account the electron-ion interactions. This relaxes the norm-conserving criteria and in doing so produces a smooth and computationally efficient pseudopotential without affecting the accuracy to a significant extent. Density mixing electronic minimiser has been used for the self-consistent calculations and Broyden Fletcher Goldfarb Shanno (BFGS) geometry optimization [24] was employed to optimize the crystal structure of the $Pmn2_1$ (No. 31) space group. To perform pseudo atomic calculations, the following electronic orbitals have been used for Mo and Te respectively: Mo [$4s^2\ 4p^6\ 4d^5\ 5s^1$], Te [$5s^2\ 5p^4$]. Periodic boundary conditions are used to determine the total energies of each cell and the trial wave functions are expanded in plane wave basis. k-point sampling within the Brillouin zone for $MoTe_2$ was carried out with 7x5x3 special points in the Monkhorst-pack grid scheme [25]. Plane wave basis set cut-off energy was taken as 350 eV. This ensures satisfactory level of convergence of the energy during cell volume calculations. Geometry optimization was carried out using a self-consistent convergence threshold of $10^{-6}$ eV atom$^{-1}$ for the energy, 0.03 eV Å$^{-1}$ for the maximum force, 0.05 GPa for maximum stress and $10^{-3}$ Å for maximum displacement.

At this point it is worth mentioning that the surface electronic states are one of the prime features of topological systems which arise from spin orbit coupling (SOC). In this investigation we have concentrated on the bulk electronic, optical and mechanical properties of orthorhombic $MoTe_2$. Variety of prior studies on diverse class of materials showed that [26–30], as far as optimization of the cell structure, bulk elastic constants, bonding and bulk optical properties are concerned, SOC only has a minimal effect. Within the bulk electronic band structure, SOC mainly reveals itself in split bands with splitting energy of the order of tens of meV. Therefore, we have not included SOC for our bulk elastic, electronic and optoelectronic calculations.

The crystal structure of $MoTe_2$ is schematically shown in Fig. 1. Here $MoTe_2$ crystallizes into the orthorhombic $T_d$ phase. We used experimental lattice parameters and atomic positions for our simulation [13]. There are 4 Mo atoms and 8 Te atoms in the unit cell. All the atoms are located at the *2a* sites with reduced positions of Mo: (0.0, 0.607140, 0.500455), (0.0, 0.031794, 0.014735); and Te: (0.0, 0.863635, 0.661450), (0.0, 0.640177, 0.112576), (0.0, 0.292018, 0.854060), and (0.0, 0.215661, 0.401925). The unit cell of $MoTe_2$

contains 12 atoms or four formula units with Te atoms forming a distorted octahedron around Mo atoms.

The optimized lattice parameters and cell volume are listed in Table I along with the previously reported theoretical and experimental values. From this table it can be seen that the calculated structural parameters are in good agreement with the corresponding experimental values. The calculated cell volume is 3-4% smaller than the experimental values. Application of GGA, on the other hand, grossly overestimates the cell volume. It is perhaps worth pointing out that the XRD results were obtained at elevated temperatures. The theoretically optimised geometry, on the other hand, corresponds to the ground state at a temperature of 0 K. Therefore, the theoretical values are expected to be somewhat lower than the experimental ones because thermal expansion is neglected and the use of LDA underestimates the lattice constants. Furthermore, the calculated elastic constants, moduli and compressibility showed that $MoTe_2$ is a soft compound with relatively weak interatomic bonding. This implies that the effect of temperature on the cell volume should be significant.

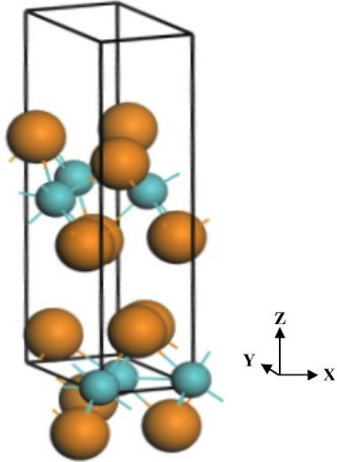

**FIG. 1.** Crystal structure of $T_d$-$MoTe_2$. The blue balls represent Mo atoms and the orange balls represent Te atoms.

**TABLE I:** The structural parameters for $MoTe_2$ obtained from the LDA calculations. The lattice parameters *a*, *b*, and *c* are in Å and the unit cell volume (*V*) is in Å$^3$.

|  | a | b | c | V |
|---|---|---|---|---|
| This work | 3.458 | 6.297 | 13.294 | 289.48 |
| Theoretical [6] | 3.469 | 6.33 | 13.83 | 303.69 |
| Experimental [8] | 3.458 | 6.304 | 13.859 | 302.11 |
| Experimental [13] | 3.477 | 6.335 | 13.889 | 305.97 |
| Experimental [14] | 3.465 | 6.307 | 13.843 | 302.52 |

## III. Results and analysis

A. Elastic properties

Elastic properties of a solid provide link between the mechanical and dynamical behaviour of the crystal. Elastic constants are calculated by the 'stress-strain' method as contained in the CASTEP program. The orthorhombic phase of MoTe$_2$ has, by symmetry arguments, nine independent single crystal elastic constants given by $C_{11}$, $C_{22}$, $C_{33}$, $C_{44}$, $C_{55}$, $C_{66}$, $C_{12}$, $C_{13}$, and $C_{23}$. We have tabulated these in Table II with the elastic compliance constants obtained from LDA calculation. The elastic constants obtained by employing GGA results in significantly lower values of $C_{ij}$. This is expected since for MoTe$_2$, GGA underestimates the atomic bonding strengths which in turn underestimate all the single crystal elastic constants. Elastic properties have been previously studied by Fan *et al.* for semiconducting MoTe$_2$ [31]. But for semi-metallic $T_d$-MoTe$_2$ these properties are explored for the first time in this study.

The elastic constants $C_{11}$, $C_{22}$, and $C_{33}$ measure the ability of the crystal to resist the applied mechanical stress along the crystallographic *a*, *b* and *c* directions respectively. It is found that $C_{33}$ is smaller than $C_{11}$ and $C_{22}$ which implies that the structure is more compressible in the *c*-direction. This reflects the layered feature of the compound. Bonding within the *ab*-plane is stronger than those extending in the out-of-plane directions. Elastic constants $C_{44}$, $C_{55}$, and $C_{66}$ determine the response of the crystal to shear. These elastic constants are particularly useful because the mechanical failure modes of crystalline solids are often controlled by shearing strain, rather than the uniaxial strains. $C_{44}$ signifies the indentation hardness of materials. The small value of $C_{44}$ indicates the material's inability of resisting the shear deformation in (100) plane. $C_{44} > C_{66}$ implies that the [100] (010) shear should be easier than the [100] (001) shear for MoTe$_2$. The off-diagonal shear components of the elastic constants $C_{12}$, $C_{13}$, and $C_{23}$ are due to the resistance to volume conserving orthorhombic distortion. The lowest value for $C_{13}$ reflects the observation of large difference between $C_{11}$ and $C_{33}$ in magnitude. It combines a uniaxial strain along the crystallographic *c*-direction to a functional stress component along the crystallographic *a*-direction. The elastic constants can be used to investigate the mechanical stability of crystal systems [32]. The modified necessary and sufficient Born criteria [33] for an orthorhombic system read:

$$C_{11} > 0; \; C_{11}C_{22} > C_{12}^2$$

$$C_{11}C_{22}C_{33} + 2C_{12}C_{13}C_{23} - C_{11}C_{23}^2 - C_{22}C_{13}^2 - C_{33}C_{12}^2 > 0 \quad (1)$$

$$C_{44} > 0; \; C_{55} > 0; \; C_{66} > 0$$

All these inequalities are satisfied by the elastic constants of MoTe$_2$ implying that the system under study is mechanically stable.

**TABLE II:** The single crystal elastic constants ($C_{ij}$ in GPa) and elastic compliance constants ($S_{ij}$ in 1/GPa) for MoTe$_2$.

| ij | C | S |
|---|---|---|
| 11 | 127.474 | 0.0094 |
| 22 | 142.302 | 0.0090 |
| 33 | 58.043 | 0.0203 |
| 44 | 24.273 | 0.0412 |
| 55 | 55.159 | 0.0181 |
| 66 | 62.273 | 0.0161 |
| 12 | 52.003 | -0.0030 |
| 13 | 22.590 | -0.0020 |
| 23 | 33.090 | -0.0040 |

The polycrystalline elastic moduli can be calculated from the single crystal elastic constants and compliances [34]. The calculated polycrystalline bulk modulus ($B$), shear modulus ($G$), Pugh's ratio ($B/G$), Young's modulus ($E$), Poisson's ratio ($v$), and machinability index ($\mu_M$) are all listed in Table III. It should be noted that the Voigt approximation assumes a continuous strain distribution but permits discontinuity in stress [35]. Consequently the actual stresses among grains are not totally balanced. This approximation provides with an upper bound of the polycrystalline elastic moduli. The Reuss approximation, alternatively, assumes continuous stress with discontinuous strain distribution inside the grains [36]. In this situation the deformed grains are not smoothly fitted with one another, giving rise to the lower bound of the polycrystalline elastic moduli. The Hill's approximation uses the arithmetic average of these two limits and closely represents the real situation in the polycrystalline solids through proper energy considerations [37]. For an isotropic solid both the Young's modulus and Poisson's ratio are related to the bulk modulus and shear modulus. Compared to many other binary and ternary metallic compounds [26,38–42], the elastic moduli of MoTe$_2$ are small, indicating its soft nature. Since $B > G$, the mechanical failure in MoTe$_2$ can be controlled by the applied shear component. The bulk modulus measures the resistance to a volume change due to isotropic applied pressure and the shear modulus characterizes the resistance to plastic deformation of a compound. A high (> 1.75) value of the Pugh's ratio ($B/G$) is associated with ductility, whereas a low (< 1.75) value corresponds to brittleness [43]. In addition, $B/G$ reflects the hardness of a material. The smaller the ratio ($B/G$), the harder the material. Thus, MoTe$_2$ is expected to show brittle characteristics. The Young's modulus measures the stiffness of solid compounds and estimates the resistance against longitudinal stress. The small value of $E$ indicates that MoTe$_2$ cannot withstand large tensile stress. Poisson's ratio measures the stability of a crystal against shear. Relatively small value of the Poisson's ratio for MoTe$_2$ indicates its stability against shear. Poisson's ratio can also predict the failure mode of solids with the critical value of 0.26 [44]. If $v$ is less (greater) than 0.26, the material is brittle (ductile). Therefore, $v$ indicates that MoTe$_2$ is brittle in nature. For solids where central force interaction dominates, $v$ resides within 0.25 and 0.50, and for non-central force solids, $v$ lies outside this range [45].

In a pure covalent compound, the Poisson's ratio is around 0.10 and for metallic bonding, the value is around 0.33. This implies that a mixture of metallic and covalent bonding exist in MoTe$_2$. The bulk modulus to $C_{44}$ ratio is defined as the machinability index [46]. This is an important parameter in the field of materials engineering. A high value of $\mu_M$ for MoTe$_2$ corresponds to good and easy machinability.

**TABLE III:** The isotropic bulk modulus ($B$ in GPa) and shear modulus ($G$ in GPa) for polycrystalline MoTe$_2$ obtained from the single crystal elastic constants using Voigt, Reuss and Hill's approximations. The Pugh ratio ($B/G$), Young's modulus ($E$ in GPa), Poisson's ratio ($v$), and the machinability index ($\mu_M$) are estimated from Hills approximation.

| $B_R$ | $B_V$ | $B_H$ | $G_R$ | $G_V$ | $G_H$ | $B/G$ | $E$ | $v$ | $\mu_M$ |
|---|---|---|---|---|---|---|---|---|---|
| 48.309 | 60.354 | 54.332 | 35.971 | 43.017 | 39.494 | 1.38 | 95.373 | 0.207 | 2.238 |

The directional bulk moduli for the single crystal has been calculated [34] from the elastic constants and are presented in Table IV. $B_{relax}$ is the single crystal isotropic bulk modulus which is almost the same as the one obtained from Reuss approximation. $B_{unrelax}$ gives the upper bound to the bulk modulus if $\alpha = \beta = 1$ is set and its value is the same as the one obtained from Vogit approximation. $\alpha$ and $\beta$ are the relative change of the $b$ and $c$ axis as a function of the deformation of the $a$ axis. The linear bulk modulus along the crystallographic axes can also be obtained from the pressure gradient. The small value of $B_c$ indicates that the compound under investigation is highly compressible when stress is applied along the $c$-direction. The varied values of $B_a$, $B_b$, $B_c$ imply that the material possesses significant bonding anisotropy. The anisotropy is strongest along the $c$-direction with respect to the one within the $ab$-plane.

**TABLE IV:** The bulk modulus ($B_{relax}$ in GPa) and its upper bound ($B_{unrelax}$ in GPa), bulk modulus along the orthorhombic crystallographic axes $a$, $b$, $c$ ($B_a$, $B_b$, $B_c$) and $\alpha$, $\beta$ for MoTe$_2$.

| $B_{relax}$ | $B_{unrelax}$ | $B_a$ | $B_b$ | $B_c$ | $\alpha$ | $\beta$ |
|---|---|---|---|---|---|---|
| 47.974 | 60.354 | 223.966 | 483.623 | 69.871 | 0.463 | 3.205 |

Study of elastic anisotropy is useful for materials design, particularly for compounds with layered structure. All the calculated elastic anisotropy factors are listed in Table V. These anisotropy factors were calculated using previously developed and widely employed formalisms [34,47,48]. The shear anisotropic factors measure the degree of anisotropy in the bonding strength for atoms in different crystal planes. $A_i = 1$ (i = 1, 2, 3) implies isotropic behaviour while deviation from unity implies elastic anisotropy. $A_1$ is the shear anisotropic factor for the {100} shear planes between the <011> and <010> directions, $A_2$ is for the {010} shear planes between the <101> and <001> directions, and $A_3$ is the factor for the

{001} shear planes between the <110> and <100> directions. The compressibility anisotropies of the bulk modulus along the *a* axis and *c* axis with respect to the *b* axis are $A_{Ba}$ and $A_{Bc}$, respectively which arises from the anisotropy of the linear bulk moduli. $A_B$ and $A_G$ are the percentage anisotropy in compressibility and shear respectively. These two factors allocate a zero value for a totally isotropic crystal and a value of 1 for the largest possible anisotropic crystal. All the anisotropy indices for $MoTe_2$ point to the anisotropic elastic property of the compound.

**TABLE V:** The shear anisotropic factors $A_1$, $A_2$, $A_3$, and $A_G$ (in %), $A_B$ (in %) and compressibility anisotropy factors $A_{Ba}$, $A_{Bc}$ for $MoTe_2$.

| $A_1$ | $A_2$ | $A_3$ | $A_G$ | $A_B$ | $A_{Ba}$ | $A_{Bc}$ |
|---|---|---|---|---|---|---|
| 0.692 | 1.645 | 1.503 | 0.089 | 0.111 | 0.463 | 0.145 |

B. Debye temperature

Debye temperature ($\theta_D$) is related to many physical properties of solids, such as specific heat, elastic constants, and melting temperature. At low temperatures the vibrational excitations arise only from acoustic modes. Hence, at low temperatures, $\theta_D$ calculated from the elastic constants is identical to that determined from specific heat measurements. $\theta_D$ sets the characteristic boson energy scale in Cooper pairing for the phonons involved in conventional superconductors. It is also related with phonon thermal conductivity and charge transport of crystalline solids. In this study, the Debye temperature has been estimated from the averaged sound velocity and crystal density ($\rho$) [49]. The longitudinal ($v_l$) and transverse ($v_t$) sound velocities of the polycrystalline material are obtained using the shear ($G$) and the bulk ($B$) modulus [34]. The calculated values are listed in Table VI. The low value of $\theta_D$ indicates the softness of $MoTe_2$ as expected from the calculated elastic moduli.

**TABLE VI:** The density ($\rho$ in g/cm$^3$), longitudinal, transverse, average elastic wave velocity ($v_l$, $v_t$, $v_m$ in m/s), and the Debye temperature ($\theta_D$ in K) from the average elastic wave velocity obtained from polycrystalline elastic modulus.

| $\rho$ | $v_l$ | $v_t$ | $v_m$ | $\theta_D$ |
|---|---|---|---|---|
| 8.054 | 3644.7 | 2214.4 | 2446.6 | 252.06 |

C. Electronic band structure and electronic energy density of states

Fig. 2 depicts the band dispersion along the high symmetry directions in the Brillouin zone (BZ) for MoTe$_2$. The results obtained from the LDA are shown here. The band dispersions obtained from GGA yield almost identical features. It is seen from Fig. 2 that both the electron and hole bands cross the Fermi level, $E_F$ (set at 0 eV) verifying its semi-metallic nature. Compared to bands running in the *ab*-plane, the bands along *c* axis (Γ-Z and X-U) are almost non-dispersive. This implies that effective mass of the charge carriers are high in this direction and as a consequence, anisotropy in charge transport within and out of the *ab*-plane should be observed. Along the Z-T direction in the k-space a moderately dispersive electron band crosses the Fermi level. This band has a Dirac cone like shape indicating the topological signature of $T_d$-MoTe$_2$.

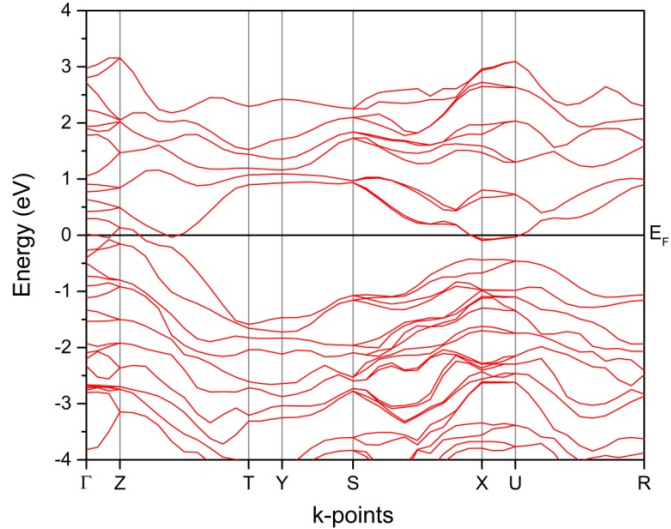

**FIG. 2.** The band structure of $T_d$-MoTe$_2$ along the high symmetry directions in the BZ.

The total density of states (TDOS) and the atom-resolved partial density of states (PDOS) of MoTe$_2$, as a function of energy, ($E$-$E_F$) are shown in Fig. 3. The vertical line at 0 eV represents the Fermi level. There is a minimum known as the pseudogap which exists almost at the Fermi level indicating that the bonding is covalent in MoTe$_2$. Pseudogap usually lies in between the bonding peak and the anti-bonding peak. These peaks are within 2 eV from $E_F$. So the Fermi level can easily be tuned by chemical or mechanical means (e.g., doping or pressure) to move across these peaks. A pseudogap very close to $E_F$ is indicative of high structural stability [50]. The calculated TDOS at $E_F$ has a value of 4.55 states/eV-unit cell. This value of $N(E_F)$ agrees well with that of Kimura *et al.* [21]. Most of the contributions in the DOS come from the Mo-4d and Te-5p electronic orbitals. The hybridization of the metal d and the chalcogen p orbitals are expected to lead to formation of the covalent bond in MoTe$_2$. Above Fermi level Te-5s orbital contributes to the DOS in the conduction band.

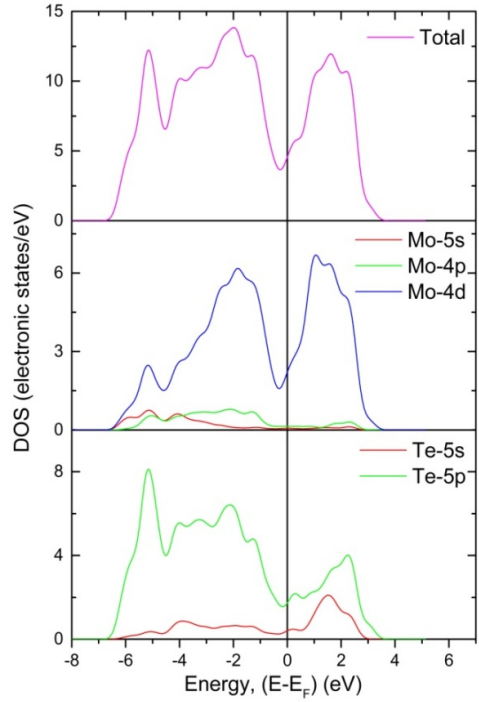

**FIG. 3.** Total and partial density of states for $T_d$-MoTe$_2$.

D. Electronic charge density distribution

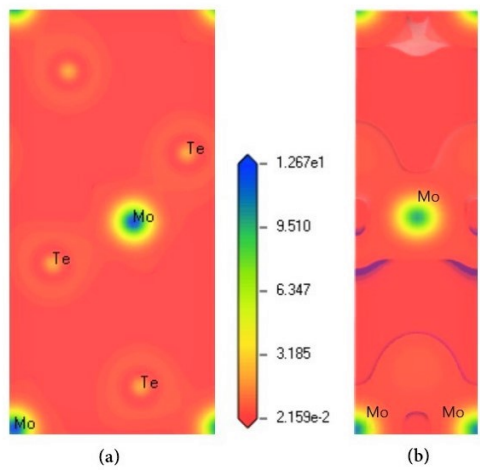

**FIG. 4.** Electronic charge density distribution map for MoTe$_2$ in the (a) (100) and (b) (010) planes.

To investigate the bonding behaviour of MoTe$_2$, the valence electron charge density distributions within the (100) and (010) planes are shown in Fig. 4. The scale in between the maps represents the total electron density. Blue indicates high electron density and red means low electron density. There is no indication of ionic bonding from the maps. Small charge accumulation between Mo and Te refers to weak covalent bonding between them. Other bonds in the compound appear to be metallic. So the bonding in MoTe$_2$ is expected to be a mixture of metallic and covalent which is entirely consistent with the elastic moduli, and Pugh's and Poisson's ratio calculations.

E. Bond population analysis

The atomic populations obtained from Mulliken population analysis (MPA) [51] and Hirshfeld population analysis (HPA) [52] are listed in Table VII. The effective valence charge (EVC) is the difference between formal ionic charge and the calculated Mulliken charge. Non-zero EVC for both the species indicate covalency in their chemical bonding. From Mulliken charge analysis we can estimate that the charge transferred from Te to Mo is 0.90e. The Mulliken population for d orbital agrees well with the earlier work by Dawson *et al.* [6]. The low value of charge spilling and Hirshfeld charge indicate that the results presented here are reliable.

The calculated bond overlap population along with the bond lengths are shown in Table VIII. Most of the overlap populations are close to zero meaning no significant interaction between the atoms. The negative values of the overlap population for Mo-Te indicate that they are anti-bonded. Bond length is a measure of the strength of chemical bonds. Our calculated values of the bond lengths indicate bonding strengths between Mo-Mo and Mo-Te are quite close and comparable. The bond lengths we obtained show very good agreement with the earlier work [6].

**TABLE VII:** Charge spilling parameter (%), orbital charges (electron), atomic Mulliken charge (electron), Hirshfeld charge (electron), and EVC (electron) of MoTe$_2$.

| Atoms | No. of ions | Charge spilling | s | p | d | Total | Mulliken charge | Hirshfeld charge | EVC |
|---|---|---|---|---|---|---|---|---|---|
| Mo | 4 | 0.25 | 2.65 | 6.97 | 5.28 | 14.91 | -0.90 | 0.04 | 4.90 |
| Te | 8 | | 1.69 | 3.85 | 0.00 | 5.55 | 0.45 | -0.02 | 1.55 |

**TABLE VIII:** Calculated bond overlap population and bond length (Å) for MoTe$_2$.

| Bond | Bond number | Population | Length (This Calc.) | Length (Calc. [6]) |
|---|---|---|---|---|
| Mo-Mo | 2 | 0.06 | 2.86 | 2.87 |
| Mo-Te | 2 | -0.06 | 2.68 | 2.68 |
| Mo-Te | 2 | -0.71 | 2.68 | 2.68 |
| Mo-Te | 2 | -0.07 | 2.69 | 2.69 |
| Mo-Te | 2 | -0.59 | 2.69 | 2.70 |
| Mo-Te | 2 | -0.14 | 2.76 | 2.80 |
| Mo-Te | 2 | -0.07 | 2.77 | 2.80 |
| Mo-Te | 2 | -0.67 | 2.79 | 2.82 |
| Mo-Te | 2 | -0.87 | 2.79 | 2.83 |

F. Optical properties

All the optical parameters presented in this section have been obtained by considering the photon induced transition probabilities between different electronic orbitals. For these calculations, a plasma frequency of 10 eV, 0.05 eV of damping in the Drude term and a Gaussian smearing of 0.5 eV were used. The calculations were performed employing LDA, because employing GGA did not change the essential features of any of the spectra. The imaginary part of the complex dielectric function, $\varepsilon_2(\omega)$, has been obtained from the matrix elements of electronic transitions between occupied and unoccupied electronic states by using the CASTEP supported formula given as,

$$\varepsilon_2(\omega) = \frac{2e^2\pi}{\Omega\varepsilon_0}\sum_{k,v,c}|\langle\psi_k^c|\hat{u}.\vec{r}|\psi_k^v\rangle|^2 \,\delta(E_k^c - E_k^v - E) \qquad (2)$$

In this expression, $\Omega$ is the volume of the unit cell, $\omega$ is the frequency of the incident electromagnetic wave (photon), $e$ is the electronic charge, $\psi_k^c$ and $\psi_k^v$ are the electronic states of electrons in the conduction and valence bands, respectively, with momentum $\hbar k$. The delta function ensures the conservation of energy and momentum during the photon absorption. The Kramers–Kronig transformation equation can be used to get the real part $\varepsilon_1(\omega)$ of the dielectric function from the corresponding imaginary part $\varepsilon_2(\omega)$, as they are causally connected. Knowing these two parts of the complex dielectric function enables one to calculate all the other optical parameters.

The energy dependent optical parameters, namely the real and the imaginary parts of the dielectric constants, $\varepsilon_1(\omega)$ and $\varepsilon_2(\omega)$, respectively, real part of refractive index $n(\omega)$, extinction coefficient $k(\omega)$, real and imaginary parts of the optical conductivity $\sigma_1(\omega)$ and $\sigma_2(\omega)$, respectively, reflectivity $R(\omega)$, the absorption coefficient $\alpha(\omega)$, and the loss function $L(\omega)$ are shown in Figs. 5. We have estimated the optical parameters for incident photon energies up to 20 eV with electric field polarization vectors along [100], [010] and [001] directions for MoTe$_2$. From Fig. 5(a), it is seen that $\varepsilon_1(\omega)$ becomes zero from below at around 16 eV and $\varepsilon_2(\omega)$ falls sharply and flattens to zero at the same energy. So 16 eV is the plasma frequency of the material above which it becomes transparent. The broad peak at 4 eV arises due to the electronic transitions between the bonding peak and the anti-bonding peak seen in the DOS spectrum [Fig. 3]. The real part of the refractive index determines the phase velocity of the electromagnetic wave in the material. The imaginary part, often termed as the extinction coefficient, in contrast, gives a measure of the attenuation as the electromagnetic wave propagates through the compound. The real part of the refractive index of $MoTe_2$ is very high in the visible region [see Fig. 5(b)] which has implications in optoelectronic device applications. Optical conductivities are finite at zero energy as can be seen from Fig. 5(c) indicating that the system under study is metallic in nature.

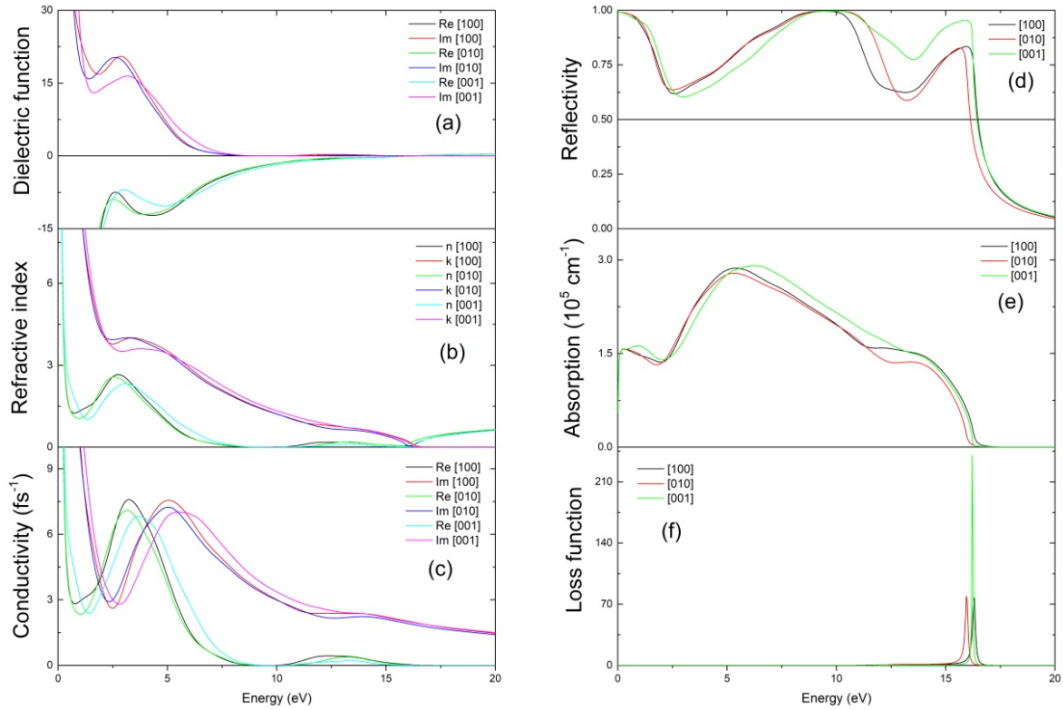

**FIG. 5.** The frequency dependent (a) dielectric function (real & imaginary), (b) refractive index (real & imaginary), (c) optical conductivity (real & imaginary), (d) reflectivity, (e) absorption coefficient, and (f) loss function of $MoTe_2$ for different polarization directions.

The reflectivity spectra shown in Fig. 5(d) fall sharply at plasma frequency. It is interesting to note that over an extended energy range up to 16 eV, $R(\omega)$ never goes below 50%, indicating that the material is good for optoelectronic device fabrication where wide band high reflectivity is required. In the infrared region, the reflectivity starts decreasing slowly from almost 100% which agrees very well with the recent experimental work by Kimura *et al.* [21]. The absorption spectra in Fig. 5(e) give a peak in the ultraviolet region and become zero above plasma frequency. The nonzero value of $\alpha(\omega)$ at 0 eV is another indication of the metallic nature of the compound. The loss peak, depicted in Fig. 5(f), is found at 16 eV, at the plasma edge as expected. All the optical constants show small anisotropy with respect to the polarization of the electric field in the *ab*-plane. The optically anisotropic feature becomes stronger when electric field lies in the *c*-direction. For example, the reflectivity increases significantly in the ultraviolet region when the electric field is along the [001] direction.

**IV. Conclusions**

To summarize, based on the DFT calculations, we have investigated in detail the mechanical and optical properties of $T_d$-MoTe$_2$. The anisotropic nature of the material is seen from the elastic, electronic and optical properties. The high value of machinability index implies possible application in the industrial sector. The refractive index is found to be very high in the visible region meaning that MoTe$_2$ will be a good candidate for optical display systems. The reflectivity spectra tell us that MoTe$_2$ will be a very good reflecting material over a wide band of frequencies. The gross features of the recently measured experimental optical conductivity and reflectivity [21] agree well with the theoretical result presented in this work. The elastic constants calculated in this study calls for experimental confirmation. Detailed study of elastic constants and moduli, as well as the Debye temperature, reveal soft nature of $T_d$-MoTe$_2$. Bonding along the *c*-direction is particularly weak. The Fermi level of the compound lies at the pseudogap minimum, in between bonding and anti-bonding peaks in the DOS. This, together with soft nature of the crystal structure, indicates that pressure will have a drastic effect on the electronic state of MoTe$_2$. As mentioned in the introduction, MoTe$_2$ exhibits superconductivity at 8.2 K under an applied pressure of 11.7 GPa. We believe that this is primarily due to pressure induced shift of the Fermi level to a peak in the electronic energy density of states. The electron phonon coupling constant is directly proportional to the density of states at the Fermi level, $N(E_F)$. Therefore, shift of the Fermi level to a peak in the DOS is expected to enhance the superconducting transition temperature to a great extent. This also suggests that, instead of applying external pressure, chemical pressure via suitable doping can also induce superconductivity in MoTe$_2$ at relatively high temperature. We hope our results will inspire both theorists and experimentalists to study this interesting Weyl semimetal in greater detail in near future.

# References

[1] H. Weyl, Proc. Natl. Acad. Sci. **15**, 323 (1929).

[2] B. Q. Lv, H. M. Weng, B. B. Fu, X. P. Wang, H. Miao, J. Ma, P. Richard, X. C. Huang, L. X. Zhao, G. F. Chen, Z. Fang, X. Dai, T. Qian, and H. Ding, Phys. Rev. X **5**, 31013 (2015).

[3] S. Y. Xu, I. Belopolski, N. Alidoust, M. Neupane, G. Bian, C. Zhang, R. Sankar, G. Chang, Z. Yuan, C. C. Lee, S. M. Huang, H. Zheng, J. Ma, D. S. Sanchez, B. K. Wang, A. Bansil, F. Chou, P. P. Shibayev, H. Lin, S. Jia, and M. Z. Hasan, Science **349**, 613 (2015).

[4] A. A. Soluyanov, D. Gresch, Z. Wang, Q. Wu, M. Troyer, X. Dai, and B. A. Bernevig, Nature **527**, 495 (2015).

[5] T. Zandt, H. Dwelk, C. Janowitz, and R. Manzke, J. Alloys Compd. **442**, 216 (2007).

[6] W. G. Dawson and D. W. Bullett, J. Phys. C Solid State Phys. **20**, 6159 (1987).

[7] Y. Sun, S. C. Wu, M. N. Ali, C. Felser, and B. Yan, Phys. Rev. B - Condens. Matter Mater. Phys. **92**, 161107 (2015).

[8] Z. Wang, D. Gresch, A. A. Soluyanov, W. Xie, S. Kushwaha, X. Dai, M. Troyer, R. J. Cava, and B. A. Bernevig, Phys. Rev. Lett. **117**, 56805 (2016).

[9] L. Huang, T. M. McCormick, M. Ochi, Z. Zhao, M. T. Suzuki, R. Arita, Y. Wu, D. Mou, H. Cao, J. Yan, N. Trivedi, and A. Kaminski, Nat. Mater. **15**, 1155 (2016).

[10] K. Deng, G. Wan, P. Deng, K. Zhang, S. Ding, E. Wang, M. Yan, H. Huang, H. Zhang, Z. Xu, J. Denlinger, A. Fedorov, H. Yang, W. Duan, H. Yao, Y. Wu, S. Fan, H. Zhang, X. Chen, and S. Zhou, Nat. Phys. **12**, 1105 (2016).

[11] J. Jiang, Z. K. Liu, Y. Sun, H. F. Yang, C. R. Rajamathi, Y. P. Qi, L. X. Yang, C. Chen, H. Peng, C. C. Hwang, S. Z. Sun, S. K. Mo, I. Vobornik, J. Fujii, S. S. P. Parkin, C. Felser, B. H. Yan, and Y. L. Chen, Nat. Commun. **8**, 13973 (2017).

[12] B. Yan and C. Felser, Annu. Rev. Condens. Matter Phys. **8**, 337 (2017).

[13] Y. Qi, P. G. Naumov, M. N. Ali, C. R. Rajamathi, W. Schnelle, O. Barkalov, M. Hanfland, S. C. Wu, C. Shekhar, Y. Sun, V. Süß, M. Schmidt, U. Schwarz, E. Pippel, P. Werner, R. Hillebrand, T. Förster, E. Kampert, S. Parkin, R. J. Cava, C. Felser, B. Yan, and S. A. Medvedev, Nat. Commun. **7**, 11038 (2016).

[14] C. Heikes, I. L. Liu, T. Metz, C. Eckberg, P. Neves, Y. Wu, L. Hung, P. Piccoli, H. Cao, J. Leao, J. Paglione, T. Yildirim, N. P. Butch, and W. Ratcliff, Phys. Rev. Mater. **2**, 74202 (2018).

[15] H. Takahashi, T. Akiba, K. Imura, T. Shiino, K. Deguchi, N. K. Sato, H. Sakai, M. S.

Bahramy, and S. Ishiwata, Phys. Rev. B **95**, 100501 (2017).

[16] Y. Zhang, Y. Sun, and B. Yan, Phys. Rev. B **97**, 41101 (2018).

[17] Y. Naidyuk, O. Kvitnitskaya, D. Bashlakov, S. Aswartham, I. Morozov, I. Chernyavskii, G. Fuchs, S. L. Drechsler, R. Hühne, K. Nielsch, B. Büchner, and D. Efremov, 2D Mater. **5**, 45014 (2018).

[18] C. Cao, X. Liu, X. Ren, X. Zeng, K. Zhang, D. Sun, S. Zhou, Y. Wu, Y. Li, and J.-H. Chen, 2D Mater. **5**, 044003 (2018).

[19] S. Lee, J. Jang, S. Il Kim, S. G. Jung, J. Kim, S. Cho, S. W. Kim, J. Y. Rhee, K. S. Park, and T. Park, Sci. Rep. **8**, 13937 (2018).

[20] D. D. Liang, Y. J. Wang, W. L. Zhen, J. Yang, S. R. Weng, X. Yan, Y. Y. Han, W. Tong, W. K. Zhu, L. Pi, and C. J. Zhang, AIP Adv. **9**, 55015 (2019).

[21] S. I. Kimura, Y. Nakajima, Z. Mita, R. Jha, R. Higashinaka, T. D. Matsuda, and Y. Aoki, Phys. Rev. B **99**, 195203 (2019).

[22] W. Kohn and L. J. Sham, Phys. Rev. **140**, A1133 (1965).

[23] S. J. Clark, M. D. Segall, C. J. Pickard, P. J. Hasnip, M. I. J. Probert, K. Refson, and M. C. Payne, Zeitschrift Fur Krist. **220**, 567 (2005).

[24] T. H. Fischer and J. Almlöf, J. Phys. Chem. **96**, 9768 (1992).

[25] H. J. Monkhorst and J. D. Pack, Phys. Rev. B **13**, 5188 (1976).

[26] M. I. Naher, F. Parvin, A. K. M. A. Islam, and S. H. Naqib, Eur. Phys. J. B **91**, (2018).

[27] M. T. Nasir, M. A. Hadi, M. A. Rayhan, M. A. Ali, M. M. Hossain, M. Roknuzzaman, S. H. Naqib, A. K. M. A. Islam, M. M. Uddin, and K. Ostrikov, Phys. Status Solidi Basic Res. **254**, 1700336 (2017).

[28] M. A. Alam, M. A. Hadi, M. T. Nasir, M. Roknuzzaman, F. Parvin, M. A. K. Zilani, A. K. M. A. Islam, and S. H. Naqib, J. Supercond. Nov. Magn. **29**, 2503 (2016).

[29] M. A. Hadi, M. A. Alam, M. Roknuzzaman, M. T. Nasir, A. K. M. A. Islam, and S. H. Naqib, Chinese Phys. B **24**, 117401 (2015).

[30] M. T. Nasir, M. A. Hadi, S. H. Naqib, F. Parvin, A. K. M. A. Islam, M. Roknuzzaman, and M. S. Ali, Int. J. Mod. Phys. B **28**, 1550022 (2014).

[31] F. Zeng, W. B. Zhang, and B. Y. Tang, Chinese Phys. B **24**, 97103 (2015).

[32] M. Born, Math. Proc. Cambridge Philos. Soc. **36**, 160 (2013).

[33] F. Mouhat and F. X. Coudert, Phys. Rev. B - Condens. Matter Mater. Phys. **90**, 224104 (2014).

[34] P. Ravindran, L. Fast, P. A. Korzhavyi, B. Johansson, J. Wills, and O. Eriksson, J. Appl. Phys. **84**, 4891 (1998).

[35] W. Voigt, *Lehrbuch Der Kristallphysik* (Teubner Leipzig, 1928).

[36] A. Reuss, ZAMM - J. Appl. Math. Mech. / Zeitschrift Für Angew. Math. Und Mech. **9**, 49 (1929).

[37] R. Hill, Proc. Phys. Soc. Sect. A **65**, 349 (1952).

[38] I. Hattabi, A. Abdiche, S. H. Naqib, and R. Khenata, Chinese J. Phys. **59**, 449 (2019).

[39] M. M. Hossain and S. H. Naqib, Mol. Phys. 1 (2019).

[40] F. Parvin and S. H. Naqib, J. Alloys Compd. **780**, 452 (2019).

[41] P. Barua, M. M. Hossain, M. A. Ali, M. M. Uddin, S. H. Naqib, and A. K. M. A. Islam, J. Alloys Compd. **770**, 523 (2019).

[42] F. Parvin and S. H. Naqib, Chinese Phys. B **26**, 106201 (2017).

[43] S. F. Pugh, London, Edinburgh, Dublin Philos. Mag. J. Sci. **45**, 823 (1954).

[44] G. N. Greaves, A. L. Greer, R. S. Lakes, and T. Rouxel, Nat. Mater. **10**, 823 (2011).

[45] O. L. Anderson and H. H. Demarest, J. Geophys. Res. **76**, 1349 (1971).

[46] Z. Sun, D. Music, R. Ahuja, and J. M. Schneider, Phys. Rev. B - Condens. Matter Mater. Phys. **71**, 193402 (2005).

[47] C. M. Kube, AIP Adv. **6**, 95209 (2016).

[48] S. I. Ranganathan and M. Ostoja-Starzewski, Phys. Rev. Lett. **101**, 55504 (2008).

[49] O. L. Anderson, J. Phys. Chem. Solids **24**, 909 (1963).

[50] J. H. Xu, T. Oguchi, and A. J. Freeman, Phys. Rev. B **35**, 6940 (1987).

[51] R. S. Mulliken, J. Chem. Phys. **23**, 1833 (1955).

[52] F. L. Hirshfeld, Theor. Chem. Acc. **44**, 129 (1977).

**Competing Interests**